\documentstyle[11pt,newpasp,twoside,epsf]{article}
\def\edcomment#1{\iffalse\marginpar{\raggedright\sl#1\/}\else\relax\fi}
\reversemarginpar

\begin{document}
\title{Ly$\alpha$ emission from GRB host galaxies}
 \author{Johan P.U. Fynbo, B. Thomsen, M.P. Egholm, M. Weidinger}
\affil{Department of Physics and Astronomy, University of \AA rhus, Ny Munkegade, 
DK-8000 \AA rhus C, Denmark}
\author{Palle M\o ller}
\affil{European Southern Onservatory, Karl Schwarzschild-Strasse 2, D-85748 Garching bei M{\"u}nchen, Germany}
\author{Jens Hjorth, Holger Pedersen, Brian L. Jensen}
\affil{Astronomical Observatory, University of Copenhagen, Juliane Maries Vej 30, DK-2100 Copenhangen \O, Denmark}
\author{Javier Gorosabel}
\affil{Instituto de Astrof\'{\i}sica de Andaluc\'{\i}a (IAA-CSIC),
P.O. Box 03004, E-18080 Granada, Spain}
\author{Michael I. Andersen}
\affil{Astrophysikalisches Institut Potsdam, An der Sternwarte 16,
D-14482 Potsdam, Germany}
\author{Stephen T. Holland}
\affil{Department of Physics, University of Notre Dame,
Notre Dame, IN 46556-5670, U.S.A.
}
\begin{abstract}
Ly$\alpha$ emission is indicative of on-going star formation in a dust-poor
environment. Ly$\alpha$ imaging is therefore a probe of the star formation 
rate and of the dust-content of Gamma-Ray Burst host 
galaxies. Both of these parameters are central to our understanding of GRB 
progenitors and of how the environments affect the propagation of afterglow 
emission out of host galaxies.
We have started a program aimed at imaging high redshift ($z>2$) host galaxies 
of GRBs at the Ly$\alpha$ resonance line from neutral hydrogen. 
Here were report the results from imaging of the fields of GRB~000301C and
GRB~000926 and outline upcoming observations of further hosts.
\end{abstract}

\section{Introduction}
The ability since 1997 to determine 
precise positions of GRB afterglows have given us a new method by which to 
locate and study galaxies in the early universe -- the host galaxies. An 
important aspect of Gamma Ray Burst selection compared to other selection 
mechanisms
is that it is not flux limited. This is obviously the case for most other 
selection methods; Lyman-Break selection (Steidel \& Hamilton 1992, Steidel 
et al. 1996) is continuum flux limited and Ly$\alpha$ selection (e.g. 
M\o ller \& Warren 1993; Cowie \& Hu 1998; Fynbo et al. 2001a) is
line flux limited.
Once an afterglow position has been determined we just need to keep 
integrating on this position until the host galaxy emerges from the noise. 
So far this approach has led to the detection of a host galaxy for all 
well localised GRBs 
(see e.g. the review by Fruchter in these proceedings) 
with the possible exception of GRB~020124 (Berger et al. 2002). Therefore,
GRB selection allows us to probe the part of the luminosity function 
currently inaccessible to other techniques. Of course GRB selection is subject 
to other selection mechanisms that are not yet completely known. The precise 
nature of these will tell us (something) about the nature of the GRB 
progenitors.
 
\section{Ly$\alpha$ emission from GRB host galaxies}

Ly$\alpha$ emission is indicative of on-going star formation in a dust-poor 
environment (Charlot \& Fall 1993, Valls-Gabaud 1993). Ly$\alpha$ imaging is 
therefore a probe of the star formation rate as well as of the dust-content of 
GRB hosts. Both of these parameters are central to our understanding of GRB 
progenitors and of how the properties of the environment affect the propagation 
of afterglow emission out of host galaxies. In the currently favored scenario
the progenitors are very massive stars (Woosley 1993 and the review in these 
proceedings). Due to their short life-times massive stars are only found in 
galaxies with on-going star 
formation.  As for the dust it is currently debated whether the non-detection 
of optical afterglows for 50--70\% of well-localised GRBs is due to dust in 
the hosts or due to intrinsic properties of the bursts
(Fynbo et al. 2001; Lazzati et al. 2002; Reichart \& Price 2002; Berger et al.
2002; De Pasquale et al. 2002). Such a bias would imply that the current sample
of GRB hosts, most of which have been located via bright optical afterglows,
would be biased against dusty hosts. From simulations there are furthermore 
indications that the ability for a super-massive star to produce a GRB is 
metallicity dependent such that low metallicities are favored (Woosley, these 
proceedings). If true, this could also imply a bias against very dusty GRB host 
galaxies. Finally, Ly$\alpha$ emission is an efficient way to probe the faint
parts of the luminosity function relevant for the majority of GRB hosts 
(Fynbo et al. 2001a) as the equivalent width of the Ly$\alpha$ line can be 
very high, several hundred \AA ngstr{\o}m in the rest frame. 

\subsection{Observations of the fields of GRB~000301C and GRB~000926}

\begin{figure}
\plotfiddle{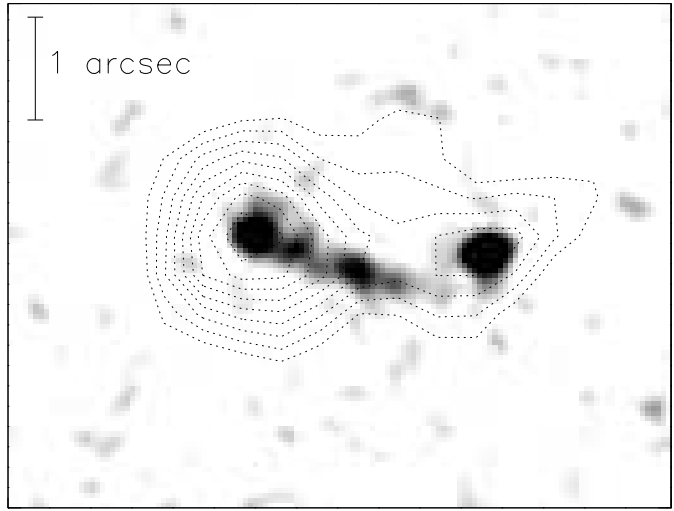}{7.5cm}{0}{60}{60}{-130}{0}
\caption{The HST/WFPC2/F606W image of the $z=2.0378$ GRB~000926 host
galaxy (Castro et al. 2001) with the Ly$\alpha$ contours over-plotted as 
dotted lines (from Fynbo et al. 2002). Most
(65\%) of the Ly$\alpha$ emission is emitted from the leftmost compact
knot. The Gamma-Ray Burst occurred in the rightmost compact knot.
}
\end{figure}

Our first study in this project consisted of deep Ly$\alpha$ narrow-band
imaging of the fields of GRB~000301C at $z=2.0404$ and GRB~000926 at 
$z=2.0378$ at the 2.56-m Nordic Optical Telescope (NOT). Both GRBs had bright 
optical afterglows.
This fortunate situation of two host galaxies with nearly identical
absorption redshifts made it possible to use a single 45\AA \ wide custom
made interference filter and search for Ly$\alpha$ emission from the host
galaxies and Ly$\alpha$ emitting galaxies in their environment.
The observations were done at the NOT in May 2001 
and the results are published in Fynbo et al. (2002) and summarized here. The 
host galaxy of GRB~000926 is an extended (more than 18 kpc), strong Ly$\alpha$ 
emitter with a rest-frame equivalent width of 71$^{+20}_{-15}$ \AA \ 
(see Fig.~1). The galaxy consists of two main components and several fainter 
knots. GRB~000926 occurred in the western component, whereas most of the 
Ly$\alpha$ luminosity (about 65\%) is emitted from the eastern component. 
From HST F606W and F814W images of the host galaxy we measure the spectral 
slopes ($f_{\lambda} \propto \lambda^{\beta}$) of the two components to 
$\beta$ = $-$2.4$\pm$0.3 (east) and $\beta$ = $-$1.4$\pm$0.2 (west). This 
implies that both components contain at most modest amounts of dust (Meurer 
et al. 1999), consistent with both the observed strong Ly$\alpha$ emission 
and the spectral energy distribution of the afterglow (Fynbo et al. 2001c).
We did not detect the host galaxy of GRB~000301C in Ly$\alpha$ emission nor
in our U and I broad-band images. This is consistent with deep STIS/HST imaging
from which the host was found to be extremely faint (R$\ga$28, Fruchter in 
these proceedings; Bloom et al.  2002). The upper limit on the Ly$\alpha$  
equivalent width is $\sim150$ \AA. Therefore, the GRB000301C host galaxy may 
also be a large equivalent width Ly$\alpha$ emitter, but a 8-m class telescope 
is required to test this. We 
detect 19 other galaxies with excess emission in the narrow filter in the two 
fields.  These galaxies are most likely Ly$\alpha$ emitting galaxies
in the environment of the host galaxies. Based on these detections we conclude
that GRB~000926 occurred in one of the strongest centres of star formation
within several Mpc, whereas GRB~000301C occurred in an intrinsically very faint
galaxy far from being the strongest centre of star formation in its galactic
environment.

\section{Perspectives}

In addition to GRB~000926 Ly$\alpha$ emission has been detected from the host 
galaxies of GRB~971214 and GRB~021004 (Kulkarni et al. 1998; Ahn 2000; 
M\o ller et al. 2002; Castro-Tirado et al. 2003). In fact, all current evidence
is consistent with the 
conjecture that high-z GRB host galaxies are Ly$\alpha$ emitters. This is 
intriguing as only $\sim$25\% of the Lyman-Break selected galaxies at the 
bright end of the high-z galaxy luminosity function are Ly$\alpha$ emitters 
(Steidel et al. 2000). This may be due to a bias against 
dust in GRB hosts, either observational or intrinsic as discussed in Sect.~2. 
To study this further we will continue the program with observations 
of more high-z GRB hosts in order to expand the sample. For now we have
time to observe the host galaxies of GRB~011211 ($z=2.140$) and GRB~020124 
($z=3.20$) with Ly$\alpha$ narrow band imaging, and in the near future the 
Swift satellite mission should make it possible to build up a larger sample. 

\acknowledgments

JPUF gratefully acknowledges support from an ESO research fellowship and
from the Carlsberg Foundation. STH acknowledges support from the NASA
LTSA grant NAG5--9364.
This work is supported by Danish Natural Science Research Council (SNF).

\end{document}